\documentclass[iop]{emulateapj}
\usepackage{multirow}
\usepackage{bm}
\usepackage{amssymb}
\usepackage{amsmath}
\usepackage{latexsym}
\usepackage{graphicx}
\usepackage{ifsym}
\usepackage{bm}
\usepackage{tabularx}

\newcommand{\Lsol}{\mbox{$L_\odot$}}

\shorttitle{New Extinction and Mass Estimates of 1RXS1609 B with MagAO}
\shortauthors{Wu et al.}
  
\begin{document}
\title{New extinction and mass estimates of the low-mass companion 1RXS 1609 B with the Magellan AO system: evidence of an inclined dust disk\footnotemark[$\ast$]$\dagger$}
\footnotetext[$\ast$]{As argued by \cite{P12}, GSC 06213--03158 B or [PZ99] J160930.3--210459 B might be better names.}
\footnotetext[$\dagger$]{This paper includes data gathered with the 6.5 m Magellan Clay Telescope at Las Campanas Observatory, Chile.}

\author{Ya-Lin Wu$^1$, 
 Laird M. Close$^1$, 
 Jared R. Males$^{1,}\footnotemark[6]$,\footnotetext[6]{NASA Sagan Fellow.}
 Travis S. Barman$^2$, 
 Katie M. Morzinski$^{1,}\footnotemark[6]$, 
 Katherine B. Follette$^3$,
 Vanessa P. Bailey$^1$,
 Timothy J. Rodigas$^{4,}\footnotemark[7]$,\footnotetext[7]{Carnegie Postdoctoral Fellow.}  
 Philip Hinz$^1$,
 Alfio Puglisi$^5$, 
 Marco Xompero$^5$, and
 Runa Briguglio$^5$}

\affil{$^1$Steward Observatory, University of Arizona, Tucson, AZ 85721, USA; yalinwu@email.arizona.edu}
\affil{$^2$Lunar and Planetary Laboratory, University of Arizona, Tucson, AZ 85721, USA}
\affil{$^3$Kavli Institute of Particle Astrophysics and Cosmology, Stanford University, 382 Via Pueblo Mall, Stanford, CA 94305, USA}
\affil{$^4$Department of Terrestrial Magnetism, Carnegie Institute of Washington, 5241 Broad Branch Road, NW, Washington, DC 20015, USA}
\affil{$^5$INAF-Osservatorio Astrofisico di Arcetri, Largo E. Fermi 5, I-50125 Firenze, Italy}
\journalinfo{}
\submitted{accepted to ApJL}

\begin{abstract} 
We used the Magellan adaptive optics system to image the 11 Myr substellar companion 1RXS 1609 B at the bluest wavelengths to date ($z'$ and $Y_s$). Comparison with synthetic spectra yields a higher temperature than previous studies of $T_\mathrm{eff}=2000\pm100~\mathrm{K}$ and significant dust extinction of $A_V=4.5^{+0.5}_{-0.7}$ mag. Mass estimates based on the DUSTY tracks gives 0.012--0.015 $M_\sun$, making the companion likely a low-mass brown dwarf surrounded by a dusty disk. Our study suggests that 1RXS 1609 B is one of the $\sim$25\% of Upper Scorpius low-mass members harboring disks, and it may have formed like a star and not a planet out at $\sim$320~AU.
\end{abstract}


\section{INTRODUCTION}
Discovery of substellar companions at hundreds of AU from their host stars in direct imaging surveys has posed challenges to classical formation mechanisms like core accretion \citep{P96} and disk instability \citep{B97}. The ultra-wide separations make the core-growing timescale exceedingly long, and observers have not yet discovered such long-lived protoplanetary disks. Some alternatives have been proposed, including gravitational scattering to the current location \citep{V09} and in situ star-like formation. 

Detecting and characterizing circumsubstellar disks can place constraints on formation mechanisms. Many substellar (close to planetary mass) companions have been suggested to host their own disks (separate from the primary's disk) based on O/IR emission lines or excess. The 1.28 $\mu$m Pa $\beta$ emission line was seen on GQ Lup B \citep{S07}, CT Cha B \citep{S08}, GSC 06214--00210 B \citep{B11}, and FW Tau C \citep{B14}. {\it Hubble Space Telescope} observations by \cite{Z14} also showed that GSC 06214--00210 B, GQ Lup B, and DH Tau B exhibit an optical excess at $\sim$0.3--0.7~$\mu$m, implying a mass accretion rate of $10^{-11}$--$10^{-9}~M_\sun \mbox{ yr}^{-1}$. \cite{K14} also found that 1RXS 1609 B, GQ Lup B, GSC 06214--00210 B, DH Tau B, and ROXs 12 B have redder $K'-L'$ colors than young field dwarfs. An unresolved 24 $\mu$m excess was also detected on GSC 06214--00210 B and 1RXS 1609 B \citep{B13}. In particular, \cite{K15} presented an ALMA Cycle 1 1.3 mm detection of dust continuum emission associated with FW Tau C and derived a dust mass of 1--2 $M_\oplus$. \cite{C15} further detected CO (2--1) and showed that gas is also present in this disk. Although FW Tau C could be in the brown dwarf regime ($M\sim10\pm4$~$M_\mathrm{jup}$; \citealt{K14}), these observations may represent the first direct detection of a disk around a planetary mass object.

Recently, using the 6.5m Magellan adaptive optics system (MagAO; \citealt{C13}; \citealt{Males14}) we have also detected an $r'$ (0.63 $\mu$m) excess possibly due to H$\alpha$ and a dust extinction of $A_V=$ 3--4 mag for CT Cha B \citep{W15}. All of these observations suggest that circumsubstellar disks could be common around these wide young planetary mass objects. The existence of disks also favors star-like fragmentation, but argues against a scattering origin. This is because disks may be perturbed, if not destroyed, by each encounter with another massive body (\citealt{B11}; \citealt{B13}). 

Here we present MagAO $z'$ and $Y_s$ imaging of 1RXS 1609 B, a substellar companion discovered at 320 AU (projected separation) from 1RXS J160929.1--210524 in the Upper Scorpius association by \cite{L08}. The companion is widely reported as the first directly imaged exoplanet orbiting a Sun-like star. Its mass, temperature, and spectral type were determined to be 0.008--0.011~$M_\sun$, $\sim$1800~K, and $\sim$L4 (\citealt{L08}, \citeyear{L10}; \citealt{L15}). Recently, \cite{P12} revised the age for Upper Sco to be $11\pm2$~Myr, and, based on that, they derived a higher mass $14^{+2}_{-3}$~$M_\mathrm{jup}$. In this paper, we show that 1RXS 1609 B may have some dust extinction, a higher temperature, and an earlier spectral type. Therefore, it is more likely a low-mass brown dwarf slightly obscured by a dusty circumsecondary disk.

\section{OBSERVATIONS and REDUCTION}
We used MagAO, a new AO system on the 6.5 m Clay Telescope, to image 1RXS 1609 A and B at $z'~(\lambda=0.91~\mu$m; $\Delta\lambda=0.12~\mu$m) and $Y_s~(\lambda=0.98~\mu$m; $\Delta\lambda=0.09~\mu$m) on 2013 April 6 (UT) during the second commissioning run. Seeing varied between 0\farcs6 and 0\farcs8. We used the primary ($R\sim 12.4$ mag) as the guide star, and locked the AO system at 100 Karhunen--Lo\`{e}ve modes and 400 Hz. This is a very faint target given that 50\% of the light goes to the VisAO science camera and 50\% to the pyramid wavefront sensor. That is why only 100 of 378 possible modes were corrected at 400 Hz compared to the usual 1000 Hz loop speed. The resulting FWHMs were 67 and 72 mas for $z'$ and $Y_s$, respectively. For $z'$, we obtained 20 s $\times$ 193 (3860 s) and 2.27 s $\times$ 28 (63.6 s) for saturated and unsaturated data, respectively. For $Y_s$, we obtained 20 s $\times$ 126 (2520 s) unsaturated frames. We only detected 1RXS 1609 B at $z'$ but not $Y_s$, possibly due to a lower quantum efficiency which leads to a lower total throughput (0.013 versus 0.077). To calibrate our photometry, we retrieved K7V, M0V, and M1V spectral templates from the Pickles Atlas \citep{P98}, reddened them by $A_V=0.1$~mag (extinction of A; \citealt{B14}), and integrated the DENIS \citep{E97} {\it I} and our $z'$ and $Y_s$ filter curves over these templates as well as the Vega spectrum to derive $I-z'$ and $I-Y_s$ colors. Then we applied these colors to the existing DENIS $I$ measurement on the primary to obtain $z'$ and $Y_s$ photometry. We adopted 0.06 and 0.10 mag as the uncertainties for $z'$ and $Y_s$ based on comparison to F7V and M1V templates. We also observed the optical standard star LTT 3864 in the same observing run, but difficulties in dealing with Strehl ratio variation on different nights and the inherent noisiness of large aperture CCD photometry made calibrations much less precise than using the DENIS photometry. However, we found consistent results to within 10\% for the photometry for 1RXS 1609 A. This demonstrated that the primary is not a very active variable and that the DENIS $I$ band photometry is valid for the night of 2013 April 6.

In the following analysis on the $z'$ data, we selected frames with wavefront errors less than 175 nm rms, corresponding to the best two-thirds of data (129 frames). Data reduction were as detailed in \cite{W15}. We constructed a master point spread function (PSF) from the unsaturated images and performed PSF-fitting photometry. In addition, aperture photometry was done using an aperture of 1 FWHM in radius. Our final $z'$ flux is the average of both approaches, and its uncertainty included a $\sim$0.1 mag offset between them. To estimate any possible flux loss in halo subtraction, we subtracted scaled-down PSFs at the position of the companion (negative PSF injection) without removing the halo and found a consistent $\Delta z'$ with our aperture and PSF-fitting photometry. Therefore, we concluded that there was no significant flux loss when removing the radially symmetric PSF halo profile. The astrometric error budget also included image distortion \citep{W15}. Table \ref{tbl-1} summarizes the system properties.  

\begin{deluxetable}{@{}lccc@{}}
\tablecaption{Properties of 1RXS 1609 System\label{tbl-1}}
\tablenum{1}
\tablehead{
\colhead{Property} &
\colhead{Primary} &
\colhead{	} &
\colhead{Companion}
}
\startdata
Distance (pc)\tablenotemark{a,b}	& 	&	$145\pm14$& 	\\
Separation (\arcsec)\tablenotemark{c}	& 	&	$2.21\pm0.01$& 	\\
PA (\degr)\tablenotemark{c}	& 	&	$27.1\pm0.3$ & 	\\
Age (Myr)\tablenotemark{d}	& 	&	$11\pm2$ & 	\\
Spectral type	&	M$0\pm1$\tablenotemark{e}&  &	L$2\pm1$\tablenotemark{c}\\
$T_\mathrm{eff}$ (K) 	&	$4060^{+300}_{-200}$\tablenotemark{f}&	 &	$2000\pm100$\tablenotemark{c}\\
$A_V$ (mag)	&	$0.1^{+0.3}_{-0.1}$\tablenotemark{e}& &	$4.5^{+0.5}_{-0.7}$\tablenotemark{c}\\
log($L_\mathrm{bol}/L_\sun$) &	$-0.37\pm0.15$\tablenotemark{f}	& &	$-3.36\pm0.09\tablenotemark{c}$\\
Mass ($M_\sun$)	&	$0.85^{+0.20}_{-0.10}$\tablenotemark{f}& &	0.012--0.015\tablenotemark{c}	\\
$I$\tablenotemark{g}		&$10.99\pm0.03$		&				& ...\\
$z'$\tablenotemark{c}	&	$10.60\pm0.06$&	&	$21.24\pm0.15$\\
$Y_s$\tablenotemark{c}	&	$10.43\pm0.10$&	&	$>$19.46 (3$\sigma$)\\
$J$\tablenotemark{h,i}	&	$9.764\pm0.027$&	&	$17.85\pm0.12$	\\
$H$\tablenotemark{h,i}&	$9.109\pm0.023$&	&	$16.86\pm0.07$	\\
$K_s$\tablenotemark{h,i}	&	$8.891\pm0.021$&	&	$16.15\pm0.05$	\\
$3.1~\mu$m\tablenotemark{j}	&	$8.80\pm0.05$&	&	$15.65\pm0.21$	\\
$3.3~\mu$m\tablenotemark{j}	&	$8.78\pm0.05$&	&	$15.20\pm0.16$	\\
$L'$\tablenotemark{k}	&	$8.73\pm0.05$&	&	$14.8\pm0.3$	\\
$24~\mu$m (mJy)\tablenotemark{j}	&	&	$3.06\pm0.04$	&
\enddata
\tablecomments{\\${}^{\mathrm{a}}$ \cite{dZ99}. ${}^{\mathrm{b}}$ \cite{I11}. ${}^{\mathrm{c}}$ This work. ${}^{\mathrm{d}}$ \cite{P12}. ${}^{\mathrm{e}}$ \cite{B14}. ${}^{\mathrm{f}}$ \cite{L08}. ${}^{\mathrm{g}}$ DENIS (\citealt{E97}). ${}^{\mathrm{h}}$ 2MASS (\citealt{S06}). ${}^{\mathrm{i}}$ \cite{L15}. ${}^{\mathrm{j}}$ \cite{B13}. ${}^{\mathrm{k}}$ \cite{L10}.}
\end{deluxetable}

\begin{figure}
\includegraphics[angle=0,width=\columnwidth]{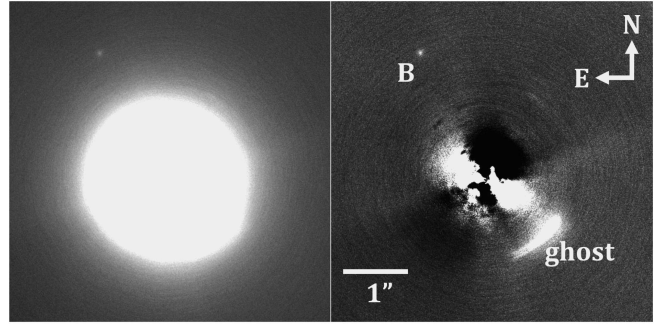}
\caption{Left: 1RXS 1609 in MagAO $z'$ filter. Right: after subtracting the radially symmetric profile of the primary star.}
\label{fig1}
\end{figure}

\begin{figure*}
\centering
\includegraphics[angle=0,width=0.418\linewidth]{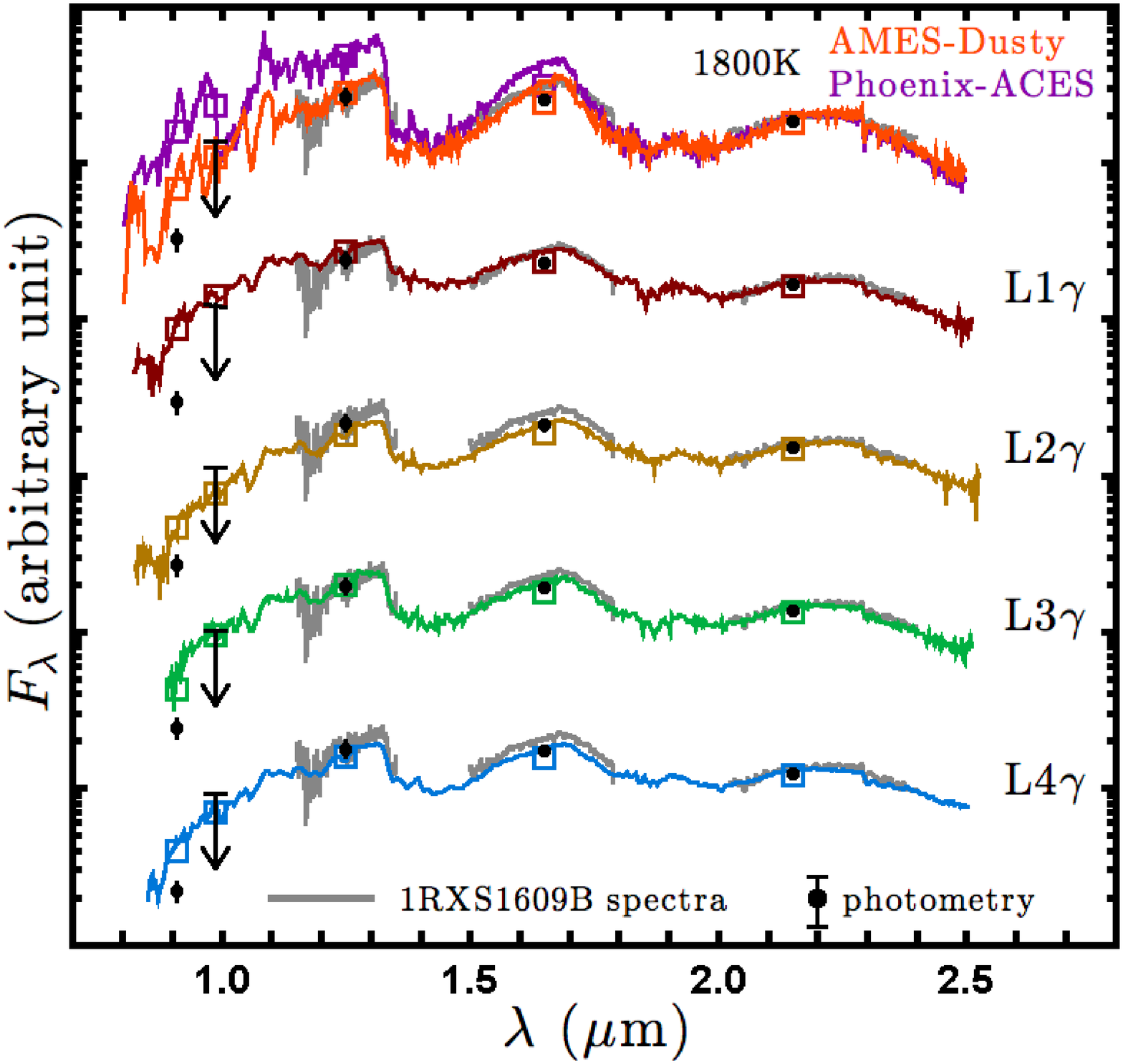}
\includegraphics[angle=0,,width=0.575\linewidth]{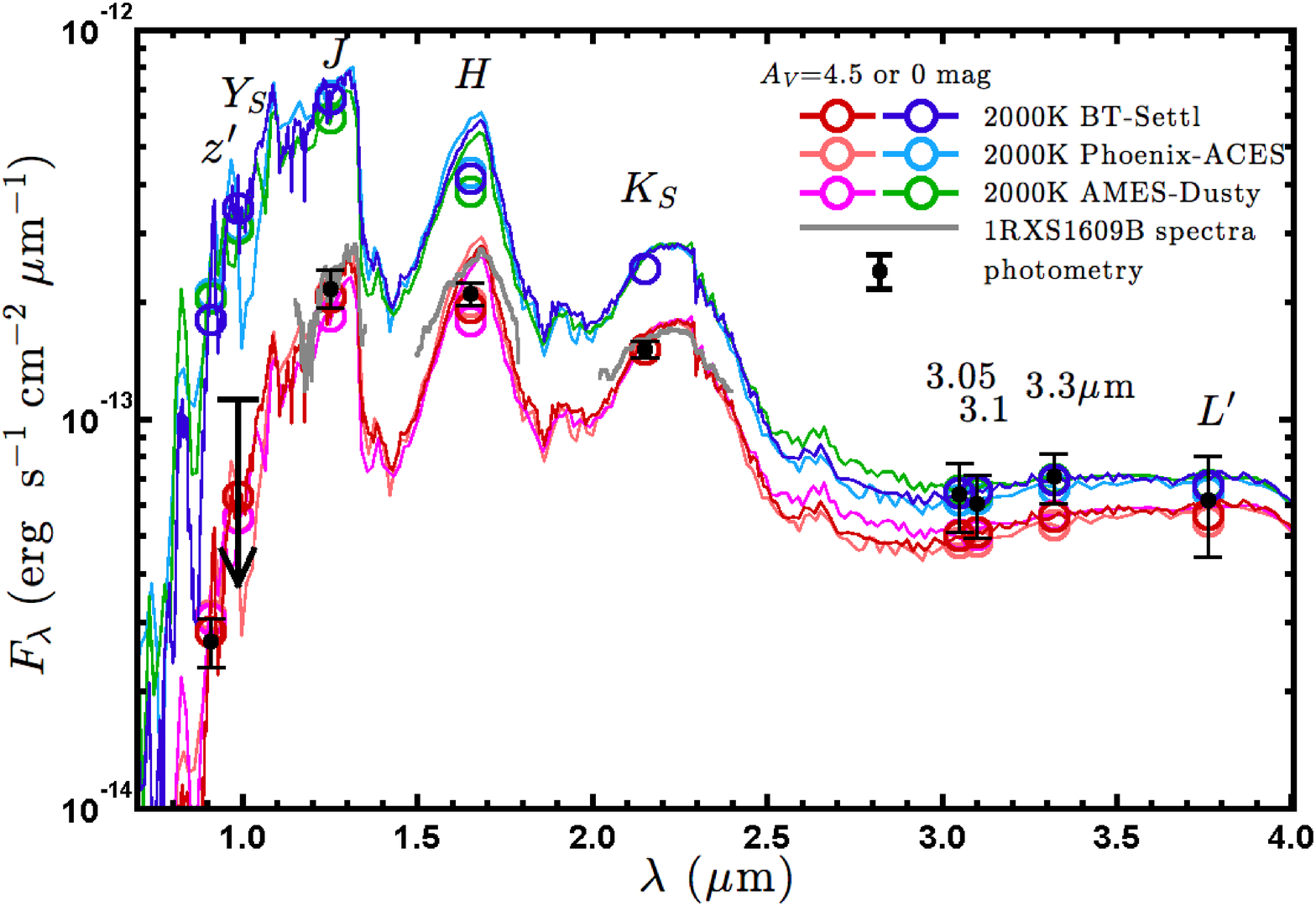}
\caption{Left: we compare the photometry and spectra of 1RXS 1609 B (\citealt{L08}, \citeyear{L10}) to the 1800~K AMES-Dusty and Phoenix-ACES models ($A_V=0$ mag) and four low-surface gravity L$\gamma$ dwarfs in \cite{AL13}. We normalize them at $K_s$. Squares represent synthetic fluxes at $z'Y_sJHK_s$, and the $Y_s$ (0.98~$\mu$m) arrow extends from $3\sigma$ down to $1\sigma$. The companion's $z'$ flux is lower than that of L$\gamma$ dwarfs and the models as well, suggesting that some dust extinction might be needed. We also note that these L$\gamma$ objects may themselves also be reddened by dust; for example, the L2$\gamma$ appears redder than the L3 and L4 counterparts. \\
Right: SED fitting with the 2000 K models. The blue curves represent what the object's spectra would be if there was no extinction. Circles denote model fluxes for filters. The $Y_s$ arrow extends from $3\sigma$ down to $1\sigma$. We need a high extinction $A_V\sim4.5$ mag (red curves) to match our observed $Y_s$ upper limit and $z'$ flux. Note that a weak ($\sim$20\%) $\sim$3~$\mu$m excess becomes apparent when the extincted SEDs are compared to the observations (3.05, 3.1, and 3.3~$\mu$m data points are slightly above the red curves).}
\label{fig2}
\end{figure*}

\begin{figure}
\centering
\includegraphics[angle=0,width=\columnwidth]{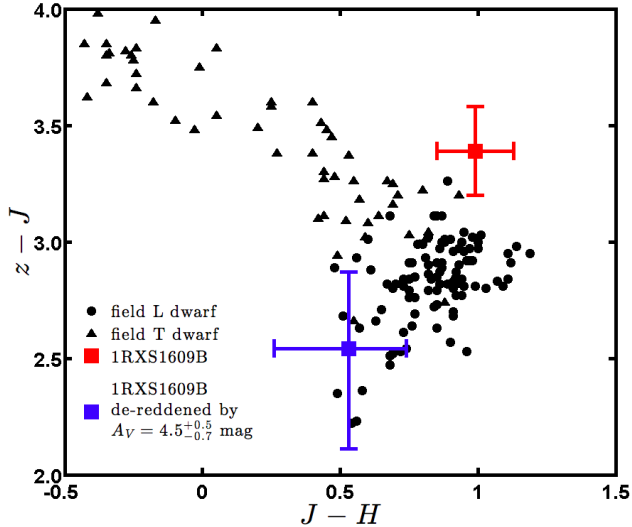}
\caption{$z-J$ vs. $J-H$ of 1RXS 1609 B and field L, T dwarfs (\citealt{G04}; \citealt{K04}; \citealt{C06}). 
Once de-reddened by $A_V\sim4.5$ mag, the color of 1RXS 1609 B is similar to that of L dwarfs.}
\label{fig3}
\end{figure}

\begin{figure}
\includegraphics[angle=0,width=\columnwidth]{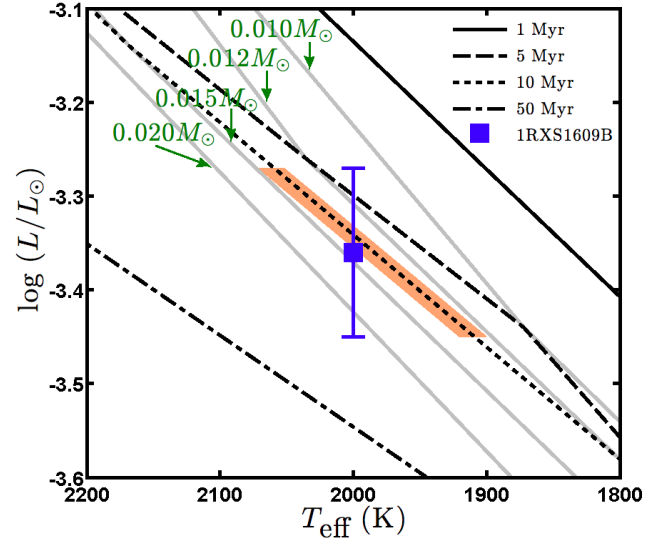}
\caption{H--R diagram with DUSTY evolutionary tracks. Iso-mass curves are shown in gray. The salmon colored polygon is the only region consistent with the observed $T_\mathrm{eff} =$ 1900--2100~K, log$(L_\mathrm{bol}/\Lsol)=-3.36\pm0.09$, and the age of Upper Sco ($11\pm2$~Myr). This yields a mass estimate of 0.012--0.015 $M_\sun$.}
\label{fig4}
\end{figure}

\section{Results}
\subsection{Properties of the Companion}
\subsubsection{Temperature and Extinction}
Figure \ref{fig1} shows MagAO $z'$ images of the system. We find a contrast of $\Delta z'=10.64\pm0.14$ mag between components. In Figure \ref{fig2}, we compare our $z'$ and $Y_s$ measurements, together with the published $JHK_s$ spectra and photometry (\citealt{L08}, \citeyear{L10}; \citealt{L15}), to the 1800~K Phoenix-ACES (\citealt{B11a}, \citeyear{B11b}), AMES-Dusty \citep{A01}, and four L$\gamma$ dwarfs\footnotemark[8]\footnotetext[8]{\cite{C09} proposed a classification scheme of $\gamma$ and $\beta$ for $\sim$10~Myr (low surface gravity) and $\sim$100~Myr (intermediate surface gravity) dwarfs, respectively.} in \cite{AL13}: 2MASS J05184616--2756457 (L1$\gamma$), 2MASS J05361998--1920396 (L2$\gamma$), 2MASS J22081363+2921215 (L3$\gamma$), and 2MASS J05012406--0010452 (L4$\gamma$). While the models and these L$\gamma$ spectra fit 1RXS 1609 B reasonably well in the near-infrared, they all seem to be too bright at $z'$ by a factor of $\sim$2--4, suggesting that some dust might be present to redden the companion. Therefore, in this study we explore other possibilities. We carry out a spectral energy distribution (SED) fitting using three atmospheric models: Phoenix-ACES, AMES-Dusty, and BT-Settl \citep{A11}. We adopt log {\it g} = 4.0 from previous analyses (\citealt{L08}, \citeyear{L10}; \citealt{L15}). Then we apply the extinction law in \cite{WD01} and \cite{D03} to redden these synthetic spectra. As in \cite{W15}, to evaluate the goodness of fit, we match these reddened models with the observed $K_s$ flux because the $K$ band is most likely dominated by photospheric emission and least affected by dust emission and extinction. In general, we find that the reddened Phoenix-ACES synthetic spectra give the best fit, followed by the BT-Settl and AMES-Dusty models. Chi-square analysis gives $T_\mathrm{eff}=2000\pm100 \mathrm{K}$, slightly higher than previous estimates. Extinction is more model dependent due to different treatments of opacity, varying from $A_V=2.7$ to 5.3 mag at this temperature range, so we compute a weighted average and adopt $A_V=4.5^{+0.5}_{-0.7}$ mag. 

Figure \ref{fig2} also shows our SED fitting with models added $A_V=0$ and 4.5 mag. We note that these fluxes are not absolute in order to avoid the $\sim$10\% distance uncertainty. We also normalize the companion's $JHK_s$ spectra (gray curves) to their apparent fluxes. The red curves are normalized at the apparent $K_s$ flux, while the blue curves represent the object's true SED when extinction is not present. Overall these reddened models fit the photometry and spectra reasonably well. The $3\sigma$ $Y_s$ upper limit also suggests that 1RXS 1609 B is likely obscured, otherwise we would have detected it. 

We also notice that the system has an unresolved 24 $\mu$m excess of 0.91 mJy \citep{B13}. Since the primary star has very little extinction, $A_V\sim0.1$ mag \citep{B14}, it is possible that most of this warm dust excess originates from the companion. As mentioned in the Introduction, \cite{K14} found that 1RXS 1609 B and a few other objects exhibit $K'-L'$ excess, which indicates that disks may be common to young substellar companions. In Figure \ref{fig2} there is evidence ($<$2~$\sigma$) of a weak $\sim$20\% 3--4~$\mu$m excess compared with our best-fit SEDs. In addition, in Figure \ref{fig3} we find that 1RXS 1609 B has $z-J$ redder than field L dwarfs (e.g., \citealt{G04}; \citealt{K04}; \citealt{C06}). Finally, the SED appears to need pure ISM-like reddening ($A_V\sim4.5$ mag), in contrast to the gray extinction usually invoked in atmospheric planetary models with thick clouds (e.g., HR 8799 b, \citealt{B11a}; 2M 1207 b, \citealt{S11}; \citealt{B11b}). All of these lines of evidence seem to suggest that 1RXS 1609 B hosts its own inclined dust disk. Really, this is not a surprising result given that $\sim$25\% of low-mass objects in Upper Sco have dusty disks \citep{LM12} and 1RXS 1609 B was the faintest, reddest object found by the discoverers' AO survey of Upper Sco \citep{L14}, so the odds are good that the faintest, reddest object might also have dust extinction as well---making 1RXS 1609 B appear redder and fainter than its intrinsic true color and luminosity. 

\subsubsection{Spectral Type, Luminosity, Mass, and Radius}
We calculate four gravity-insensitive indices $\mathrm{H}_2$O, $\mathrm{H}_2$OD, $\mathrm{H}_2$O-1, and $\mathrm{H}_2$O-2 as defined in \cite{AL13}. The average of these indices shows $\sim$L2 for $A_V$ between 3 and 6 mag, earlier than previous estimates of L$4\pm1$ (\citealt{L08}, \citeyear{L10}; \citealt{L15}). Our result is also in agreement with the recent analysis by \cite{M14}, who fit an L2$\gamma$ spectra to that of 1RXS 1609 B. Therefore, we adopt L$2\pm1$. 

With $A_V=4.5^{+0.5}_{-0.7}$ mag and $D=145\pm14$~pc, we calculate log($L_\mathrm{bol}/L_\sun)=-3.36\pm0.09$ using the bolometric correction in \cite{S14}. To validate, we integrate the de-reddened spectra in Figure \ref{fig2} and obtain $\sim-3.32$, suggesting that the bolometric correction derived from field dwarfs works reasonably well for young objects. Compared to the DUSTY evolutionary tracks (\citealt{C00}; see Figure \ref{fig4}), the companion has a mass between 0.012 and 0.015~$M_\sun$, consistent with \cite{P12} but higher than 0.008--0.011~$M_\sun$ found in other studies. Therefore, we suggest that 1RXS 1609 B likely lies above the fiducial brown dwarf/planet boundary.

Finally, we obtain $\sim$1.7 Jupiter radii using the new luminosity and temperature, consistent with the DUSTY tracks.

\subsection{Implications}
If 1RXS 1609 B harbors an inclined disk, this will imply that circumsubstellar disks could survive after 10 Myr. This is not entirely unexpected because the recent large infrared survey in the Upper Sco revealed longer disk lifetimes for low-mass stars \citep{LM12}. The survival of disks also supports in situ fragmentation for companions on wide orbits and disfavors the planet--planet scattering scenario. Since no accretion-indicating lines were detected in the NIR spectrum, ongoing accretion is either slow or non-existent. The putative disk may be largely gas depleted, precluding accretion, while still retaining sufficient dust mass at larger radii to produce the observed extinction. Future ALMA observations could definitively test the existence of this disk.

\acknowledgements
We thank the anonymous referee for helpful comments. We thank Professor David Lafreni\`{e}re for providing the companion's spectra. We thank the whole Magellan Staff for making this wonderful telescope possible. We would especially like to thank Povilas Palunas (for help over the entire MagAO commissioning run). Juan Gallardo, Patricio Jones, Emilio Cerda, Felipe Sanchez, Gabriel Martin, Maurico Navarrete, Jorge Bravo, and the whole team of technical experts helped perform many exacting tasks in a very professional manner. Glenn Eychaner, David Osip, and Frank Perez all gave expert support which was fantastic. It is a privilege to be able to commission an AO system on such a fine telescope and site. The MagAO system was developed with support from the NSF, MRI and TSIP programs. The VisAO camera was developed with help from the NSF ATI program. Y.-L.W.'s and L.M.C.'s research were supported by NSF AAG and NASA Origins of Solar Systems grants. J.R.M. is grateful for the generous support of the Phoenix ARCS Foundation. J.R.M. and K.M. were supported under contract with the California Institute of Technology, funded by NASA through the Sagan Fellowship Program. V.B. was supported in part by the NSF Graduate Research Fellowship Program (DGE-1143953). Y.-L.W. thanks Mr. Cosmos C. Yeh for some interesting discussions about academia.

\end{document}